# Pressure-induced nontrivial $Z_2$ band topology and superconductivity in transition metal chalcogenide $Ta_2Ni_3Te_5$

Haiyang Yang,[1,2] Yonghui Zhou,[1,*] Shuyang Wang,[1,2] Jing Wang,[1,2] Xuliang Chen,[1,2] Lili Zhang,[3] Chenchao Xu,[4,*] and Zhaorong Yang[1,2,5,6,*]

[1] *Anhui Key Laboratory of Condensed Matter Physics at Extreme Conditions, High Magnetic Field Laboratory, HFIPS, Chinese Academy of Sciences, Hefei 230031, China*

[2] *Science Island Branch of Graduate School, University of Science and Technology of China, Hefei 230026, China*

[3] *Shanghai Synchrotron Radiation Facility, Shanghai Advanced Research Institute, Chinese Academy of Sciences, Shanghai 201204, China*

[4] *Center for Green Research on Energy and Environmental Materials (GREEN) and International Center for Materials Nanoarchitectonics (MANA), National Institute for Materials Science (NIMS), Tsukuba, Ibaraki 305-0044, Japan*

[5] *Institutes of Physical Science and Information Technology, Anhui University, Hefei 230601, China*

[6] *Collaborative Innovation Center of Advanced Microstructures, Nanjing University, Nanjing 210093, China*

*Corresponding authors.

yhzhou@hmfl.ac.cn;

chenchaoxu.xcc@gmail.com;

zryang@issp.ac.cn

## Abstract

The unique electronic and crystal structures driven by external pressure in transition metal chalcogenides (TMCs) can host emergent quantum states. Here we report pressure-induced metallization, nontrivial $Z_2$ band topology and superconductivity in TMC $Ta_2Ni_3Te_5$. Our electrical transport measurements show that the metallization emerges at 3.3 GPa, followed by appearance of the superconductivity at $P_c = 21.3$ GPa with $T_c \sim 0.4$ K. Room-temperature synchrotron x-ray diffraction experiments demonstrate the stability of the pristine orthorhombic structure upon compression. Our first-principles calculations further reveal a topological phase transition (from $Z_2 = 0$ to $Z_2 = 1$), which occurs after $Ta_2Ni_3Te_5$ is turned into an electron-hole compensated semimetal by pressure. The pressure-induced superconductivity at $P_c$ could be attributed to the abruptly enhanced density of states at the Fermi level. These findings demonstrate that $Ta_2Ni_3Te_5$ is a new platform for realizing exotic quantum phenomena in TMCs, as well as exploring the interplay between topological property and superconductivity.

Transition metal chalcogenides (TMCs) have been intensively investigated in past decades, which not only have potential applications in electronics and optoelectronics, but also provide a platform for investigating novel quantum states, such as quantum spin Hall (QSH), high-order topology and topological superconductivity [1-5]. Recently, a new class of TMCs $Ta_2M_3Te_5$ (M = Ni, Pd) has been attracting more and more attention. Theoretical calculation suggested that the monolayer $Ta_2Pd_3Te_5$ is a QSH insulator, while the monolayer $Ta_2Ni_3Te_5$ is a trivial insulator but can be tuned into the QSH state by uniaxial strain [6]. Due to the presence of double-band inversion, it was further proposed that the monolayer $Ta_2M_3Te_5$ (M = Ni, Pd) can host a two-dimension second-order topology [7]. Consistent with the theoretical predictions, scanning tunneling microscopy measurements confirmed the existence of topological edge states in monolayer $Ta_2Pd_3Te_5$ [8]. In addition, superconductivity was induced by Ti- or W-doping in bulk $Ta_2Pd_3Te_5$, although the parent compound shows semiconducting behavior [9].

Without introducing impurities, pressure can directly modify the lattice and effectively change the physical properties of materials [10-19]. In this work, we find multiple pressure-induced quantum phase transitions in $Ta_2Ni_3Te_5$ by combining experimental measurements and theoretical calculations. At ambient pressure, the $Ta_2Ni_3Te_5$ is a semiconductor with trivial $Z_2$ band topology. Upon compression, although the pristine structure is stable up to 50.1 GPa, electrical transport measurements show that $Ta_2Ni_3Te_5$ displays a semiconductor-to-metal transition at 3.3 GPa, followed by the emergence of superconductivity at $P_c$ = 21.3 GPa. First-principles calculations suggest $Ta_2Ni_3Te_5$ is tuned to an electron-hole compensated semimetal at 2.1 GPa and undergoes a topological phase transition from $Z_2 = 0$ to $Z_2 = 1$ around 4 GPa. The nontrivial topological property is preserved at least up to 41.1 GPa, suggesting that the pressurized $Ta_2Ni_3Te_5$ possesses coexistence of nontrivial topological band structure and superconductivity.

The experimental and computational methods used in this study and supporting data under pressure are described in the Supplemental Material [20-31]. $Ta_2Ni_3Te_5$ crystallizes in a layered orthorhombic structure with space group *Pnma* (No. 62) [32], as illustrated in Fig. 1(a). The unit cell contains two $Ta_2Ni_3Te_5$ monolayers stacked along the *a* axis, coupled to each other by weak van der Waals interactions. Each monolayer comprises three atomic layers: tellurium atoms form the top and bottom layers, and the middle layer contains tantalum and nickel atoms. Figure 1(b) shows the XRD pattern of the single crystal $Ta_2Ni_3Te_5$. It is clear that only ($l$00) diffraction peaks can be detected, demonstrating that the *bc* plane is a natural cleavage facet. The inset of Fig. 1(b) exhibits a picture of as-grown $Ta_2Ni_3Te_5$ single crystals whose preferred orientation is along the *b*-axis. The crystals are as large as 2 mm and have shiny surfaces. As shown in Fig. 1(c), the energy dispersive x-ray spectroscopy gives the ratio of Ta : Ni : Te ~ 1.97 : 2.99 : 5.04, which is close to the ideal stoichiometry of Ta : Ni : Te = 2 : 3 : 5. The above characterizations demonstrate the high quality of the samples used here. Figure 1(d) presents the temperature-dependent resistivity $\rho(T)$ and dc magnetization $M(T)$ curves of $Ta_2Ni_3Te_5$ single crystals from 2 to 300 K. For the high-temperature $\rho(T)$ curve from 300 to 180 K, it can be fitted well by the Arrhenius model $\rho(T) = \rho_0 \exp(E_g/k_BT)$, where $k_B$ and $E_g$ are the Boltzmann constant and thermal activation energy, respectively. The extracted activation energy $E_g$ is ~29.8 meV. In the low temperature region, the resistivity can be better described by the Mott's variable-range hopping (VRH) model. Because $Ta_2Ni_3Te_5$ displays a paramagnetic behavior as indicated by the monotonic increase of magnetization upon cooling from 300 K, the small polaron hopping model for a magnetic semiconductor like $Mn_3Si_2Te_6$ [33] is not applicable here.

Figure 2 shows the $\rho(T)$ curves of $Ta_2Ni_3Te_5$ at various pressures up to 50.5 GPa. At 0.8 GPa, $Ta_2Ni_3Te_5$ displays a semiconducting behavior, and the overall resistivity decreases by two orders of magnitude compared to the case of ambient pressure. Upon compression to 3.3 GPa, the pressure-induced metallic conductivity can be recognized. With further increasing pressure, $Ta_2Ni_3Te_5$ remains metallic, while a tiny resistivity drop is observed around 0.5 K at 21.3 GPa. Such a drop becomes more and more pronounced at higher pressures, and the zero resistance is finally achieved at 34.2 GPa, indicating the appearance of superconductivity in $Ta_2Ni_3Te_5$, as shown in Fig. 2(b-c). Moreover, the superconductivity is robust up to 50.5 GPa, the highest pressure conducted in this study. In order to confirm the superconducting state, we further performed resistance measurements under various magnetic fields along the *a*-axis at 34.2 GPa, as shown in Fig. 2(d). By defining $T_c$ with the resistivity criterion of $\rho_{cri}$ = 90%$\rho_n$ ($\rho_n$ is the normal state resistivity), we plotted the temperature-magnetic field phase diagram in Fig. 2(e). By fitting the data to the Werthamer-Helfand-Hohenberg (WHH) model [34], the estimated upper critical field $\mu_0H_{c2}(0)$ ~ 0.61 T is much lower than the Pauli limit field of $\mu_0H_P(0) = 1.84T_c$ (~ 1.4 T), which suggests the absence of Pauli paramagnetic pair-breaking effect. According to the relationship $\mu_0H_{c2}(0)$ =

$\Phi_0/(2\pi\xi^2)$, where $\Phi_0 = 2.07 \times 10^{-15}$ Wb is the flux quantum, the coherence length $\xi$ of 232.4 Å is obtained.

To examine the structural stability of pristine $Ta_2Ni_3Te_5$ under pressure, we performed high-pressure powder XRD measurements at room temperature. As shown in Fig. 3(a), all the XRD peaks continuously shift towards higher angles without appearance of extra peaks up to 50.1 GPa. Using the Le Bail method, all XRD patterns can be indexed by the space group *Pnma*. A typical analysis of the XRD pattern at 0.7 GPa is shown at the bottom of Fig. 3(a). We extract the lattice parameters as a function of pressure, as shown in Fig. 3(b). Upon compression from 0.7 to 50.1 GPa, the parameters *a*, *b*, and *c* decreased by ~11.6%, 5.5%, and 8.4%, respectively, indicating it is much more compressible for interlayer than intralayer. As shown in Fig. 3(c), the pressure-dependent volume can be fitted by the third-order Birch-Murnaghan equation of state [35]. The fitting yields ambient pressure volume $V_0 = 907.7 \pm 3.2$ Å$^3$, bulk modulus $B_0 = 52.4 \pm 4.8$ GPa, and its first pressure derivative $B_0' = 9.6 \pm 0.8$, respectively.

To further understand the pressure evolution of electronic transport properties in $Ta_2Ni_3Te_5$, electronic band structures under pressure were calculated by using the Vienna Abinit Simulation Package (VASP) [24] with Perdew-Burke-Ernzerhof (PBE) exchange functional [25]. At ambient pressure, the band structure is fully gapped in the presence of spin-orbit coupling (SOC) (Fig. 4(a)), indicating a semiconductor ground state. The $Z_2$ index defined with the parity at the time-reversal invariant momenta (TRIM) is [0;000], exhibiting trivial topological property that is consistent with the previous study [6]. The semiconducting behavior and trivial $Z_2$ topological property were also confirmed by the calculations with modified Becke-Johnson (mBJ) exchange potential [26]. The detailed evolution of band structures under low pressure is illustrated in Fig. S2. The conduction and valence bands cross the Fermi level around 2.1 GPa, with the formation of tiny electron and hole pockets, leading to an electron-hole compensated semimetallic state.

Further increasing pressure, the direct band gap along Y-Γ gradually closes and reopens at a slightly higher pressure. Interestingly, such a gap closing-reopening transition is concomitant with the band inversion. In order to verify the topological property of $Ta_2Ni_3Te_5$, we calculated the $Z_2$ index (Table S1). Before 3.6 GPa, $Z_2$ index keeps to be [0:000]. However, after the band inversion taking place around Γ point, the $Z_2$ index becomes [1:111] at 5.3 GPa, suggesting a topological phase transition from $Z_2$ trivial to nontrivial. After extrapolating the direct band gap along Y-Γ below 10.4 GPa, the critical pressure for topological phase transition is estimated to be 4.0 GPa (Fig. S3). We further illustrate the surface state obtained from surface Green functions calculation [36] at 5.3 GPa, as shown in Fig. 5(a). In contrast to the fully gaped bulk band structure (Fig. 4(b)) at Γ point, the Dirac-type surface states appear in the spectra, demonstrating the nontrivial topological property. It is noted that the pressure-induced nontrivial topological property has also been reported in other systems, such as BiTeI, $Sb_2Se_3$, SnSe, LaSb, and γ-InSe [10-15]. For $Ta_2Ni_3Te_5$, the electron and hole pockets become larger under higher pressure (Fig. 4(c,d)), while the highest valence band and lowest conducting band remain separated. The nontrivial band topology is preserved to 41.1 GPa, which is confirmed by the $Z_2$ calculation (Table S1).

In Fig. 5(b), we show the evolution of the density of states (DOS) upon compression. A pressure-induced red shift of DOS peak around the Fermi level is observed. Around $P_c \sim 20$ GPa, at which the superconductivity emerges, the DOS peak shifts to the Fermi level. Simultaneously, the bands along T-Y-S (Fig. 4(c,d)) with weak dispersion cross the Fermi level, forming rather flat electron-type Fermi surfaces around Y point (Fig. 5(c,d)). This implies that the pressure-induced superconducting transition is probably due to the enhanced DOS at the Fermi level (Fig. S4) with large contribution from the new emerged flat electron-type surfaces around Y point.

In summary, the pressure effect on the structural and electronic properties in $Ta_2Ni_3Te_5$ is systematically investigated by combining experimental measurements and theoretical calculations, from which the pressure-temperature phase diagram of $Ta_2Ni_3Te_5$ is constructed in Fig. 6. High-pressure synchrotron XRD reveals no structural phase transition up to 50.1 GPa. Our electrical transport experiments show that the metallization of $Ta_2Ni_3Te_5$ emerges at 3.3 GPa, followed by the appearance of superconductivity at $P_c = 21.3$ GPa. The metallization can be ascribed to the formation of electron and hole pockets around the Fermi level, whereas the superconductivity originates from the abruptly enhanced density of states at the Fermi level. More interestingly, we find a topological phase transition occurs at ~4.0 GPa, featured by the band inversion around Γ point, the change of the $Z_2$ index as well as the appearance of the Dirac-like topological surface states. The nontrivial topological property maintains up to 41.1 GPa, suggesting that the superconductor $Ta_2Ni_3Te_5$ displays nontrivial band topology. These findings provide an exciting opportunity to investigate the correlation between superconductivity and topological states in transition metal chalcogenides.

## ACKNOWLEDGMENTS

The authors are grateful to Siqi Wu and Chao Cao for helpful discussion. The authors gratefully acknowledge financial support from the National Key Research and Development Program of China (Grant Nos. 2022YFA1602603 and 2018YFA0305704), the National Natural Science Foundation of China (Grant Nos. U1932152, 12174395, 12004004, and U19A2093), the Natural Science Foundation of Anhui Province (Grant Nos. 2008085QA40 and 1908085QA18), the Users with Excellence Project of Hefei Center CAS (Grant Nos. 2021HSC-UE008 and 2020HSC-UE015), the Collaborative Innovation Program of Hefei Science Center CAS (Grant No. 2020HSC-CIP014). A portion of this work was supported by the High Magnetic Field Laboratory of Anhui Province under Contract No. AHHM-FX-2020-02. Yonghui Zhou was supported by the Youth Innovation Promotion Association CAS (Grant No. 2020443). A portion of this work was supported by the High Magnetic Field Laboratory of Anhui Province. The high-pressure synchrotron X-ray diffraction experiments were performed at the beamline BL15U1, Shanghai Synchrotron Radiation Facility. The DFT calculations were performed on the supercomputers at NIMS (Numerical Materials Simulator), Institute for Solid State Physics, the University of Tokyo in Japan and the HPC center at Hangzhou Normal University in China.

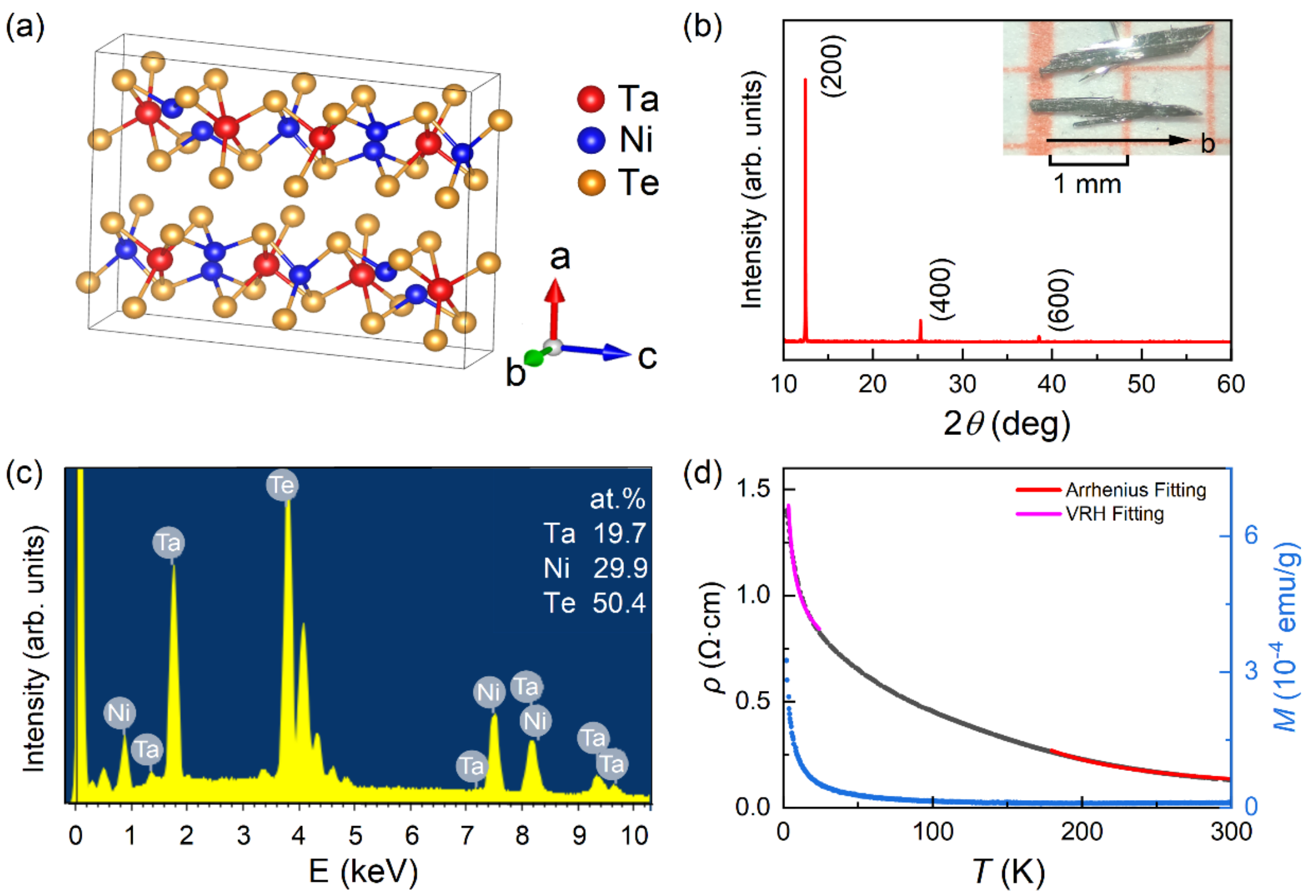

FIG. 1. (a) Schematic crystal structure of $Ta_2Ni_3Te_5$ (orthorhombic, space group *Pnma*). The red, blue, and orange spheres represent Ta, Ni, and Te, respectively. (b) XRD pattern of $Ta_2Ni_3Te_5$ single crystal. The inset shows a picture of as-grown single crystals. (c) Energy-dispersive X-ray spectroscopy. (d) (left) The $\rho(T)$ curve of $Ta_2Ni_3Te_5$ single crystal. The red and violet solid lines denote the fitting results based on the Arrhenius model and the VRH model, respectively. (right) Temperature dependence of dc magnetization $M(T)$ curve of $Ta_2Ni_3Te_5$ single crystal at $H = 1$ kOe along the *a* axis.

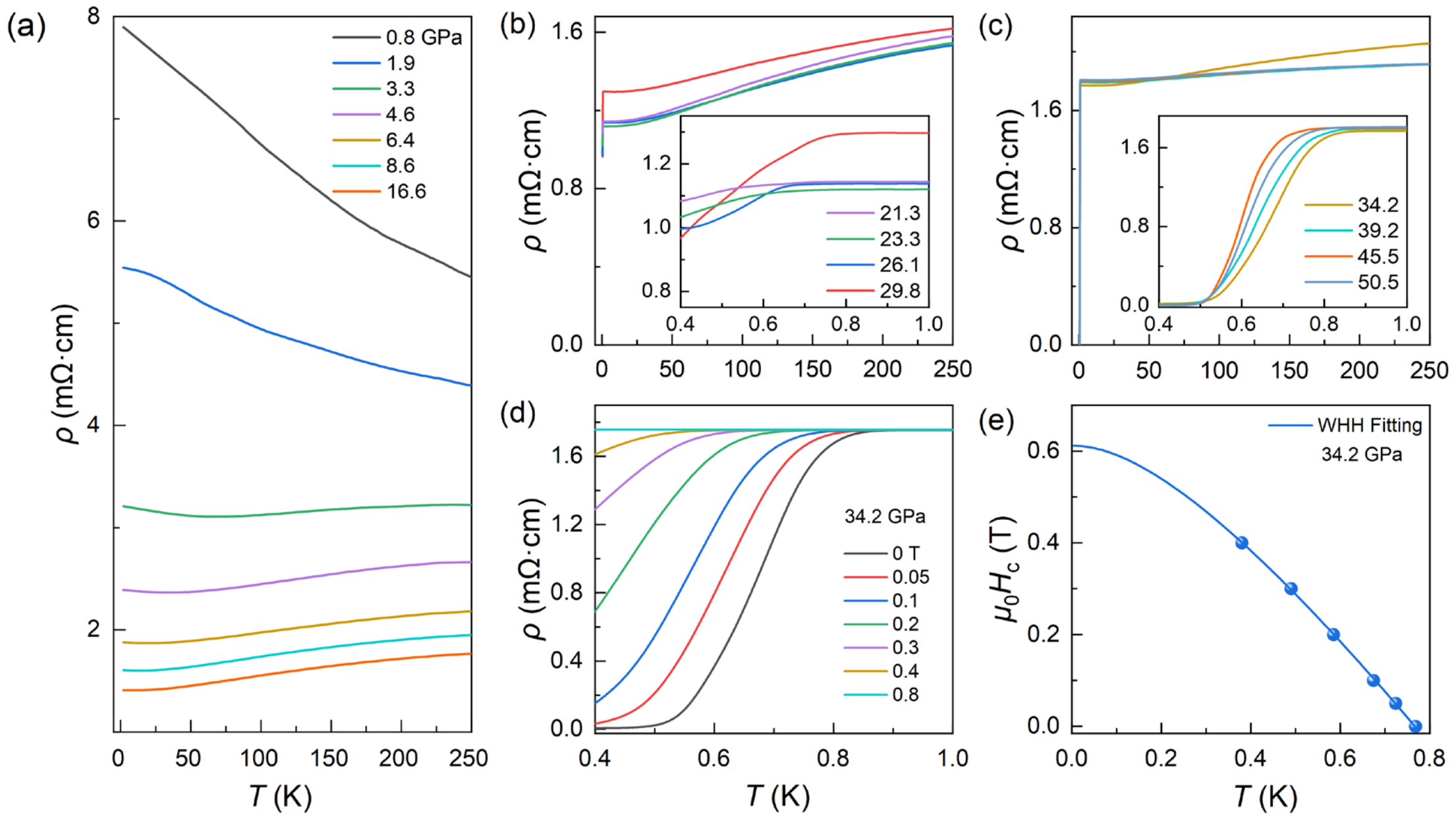

FIG. 2. (a) Temperature-resistivity curves $\rho(T)$ of $Ta_2Ni_3Te_5$ single crystal at various pressures up to 16.6 GPa. (b) and (c) show the emergence of pressure-induced superconducting transition at higher pressures: 21.3 ~ 29.8 GPa, 34.2 ~ 50.5 GPa, respectively. Inset show the curves $\rho(T)$ around superconducting transition temperatures. (d) $\rho(T)$ curves under various magnetic fields at 34.2 GPa. (e) Temperature dependence of the upper critical field $\mu_0H_{c2}$ at 34.2 GPa. The solid line represents the fitting based on the WHH model.

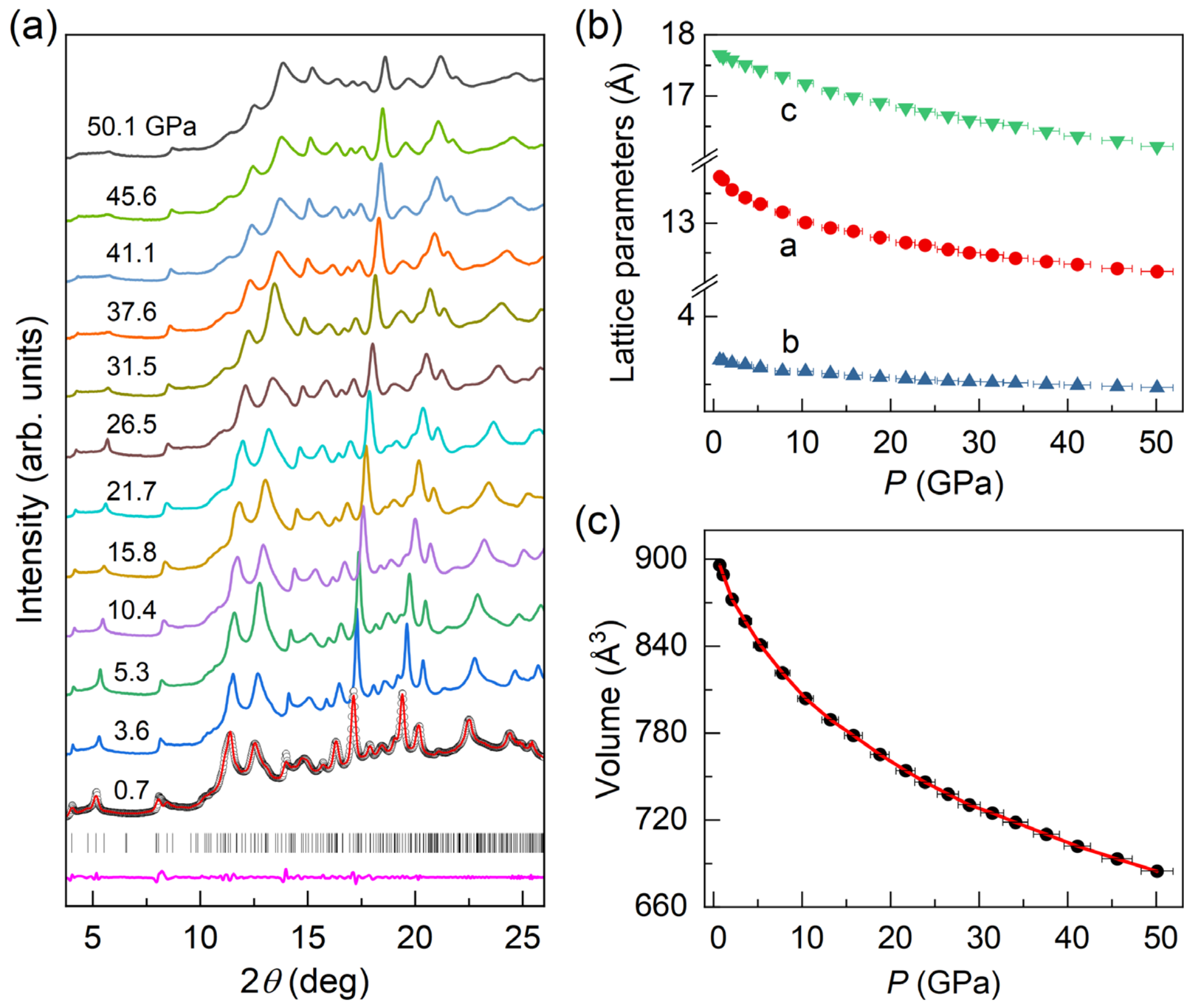

FIG. 3. (a) Pressure dependence of XRD patterns of $Ta_2Ni_3Te_5$ at room temperature ($\lambda = 0.6199$ Å). Bottom: Representative fitting of the XRD pattern at 0.7 GPa with $R_p = 0.6\%$ and $R_{wp} = 1.3\%$. The vertical bars and the violet line stand for peak positions and difference between the data and the theoretical calculation, respectively. (b) Lattice parameters *a*, *b*, and *c* as a function of pressure. (c) Volume as a function of pressure. The solid red line denotes the fitting for the pristine orthorhombic phase according to the third-order Birch-Murnaghan equation of states.

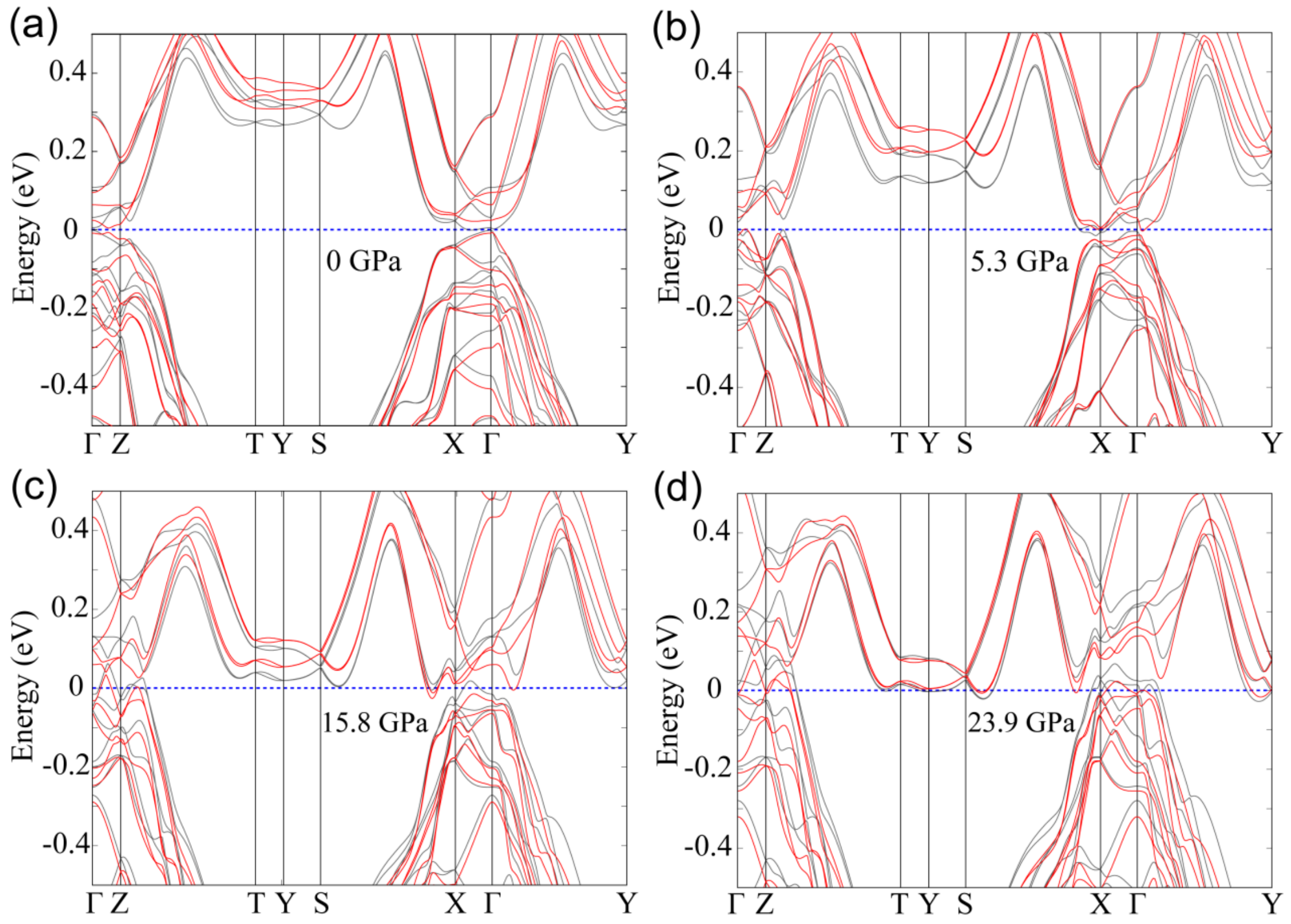

FIG. 4. (a-d) The band structure with spin-orbit coupling for ambient pressure, 5.3 GPa, 15.8 GPa and 23.9 GPa, respectively. The black line is corresponding to the PBE calculation and the red line is the corresponding mBJ calculation.

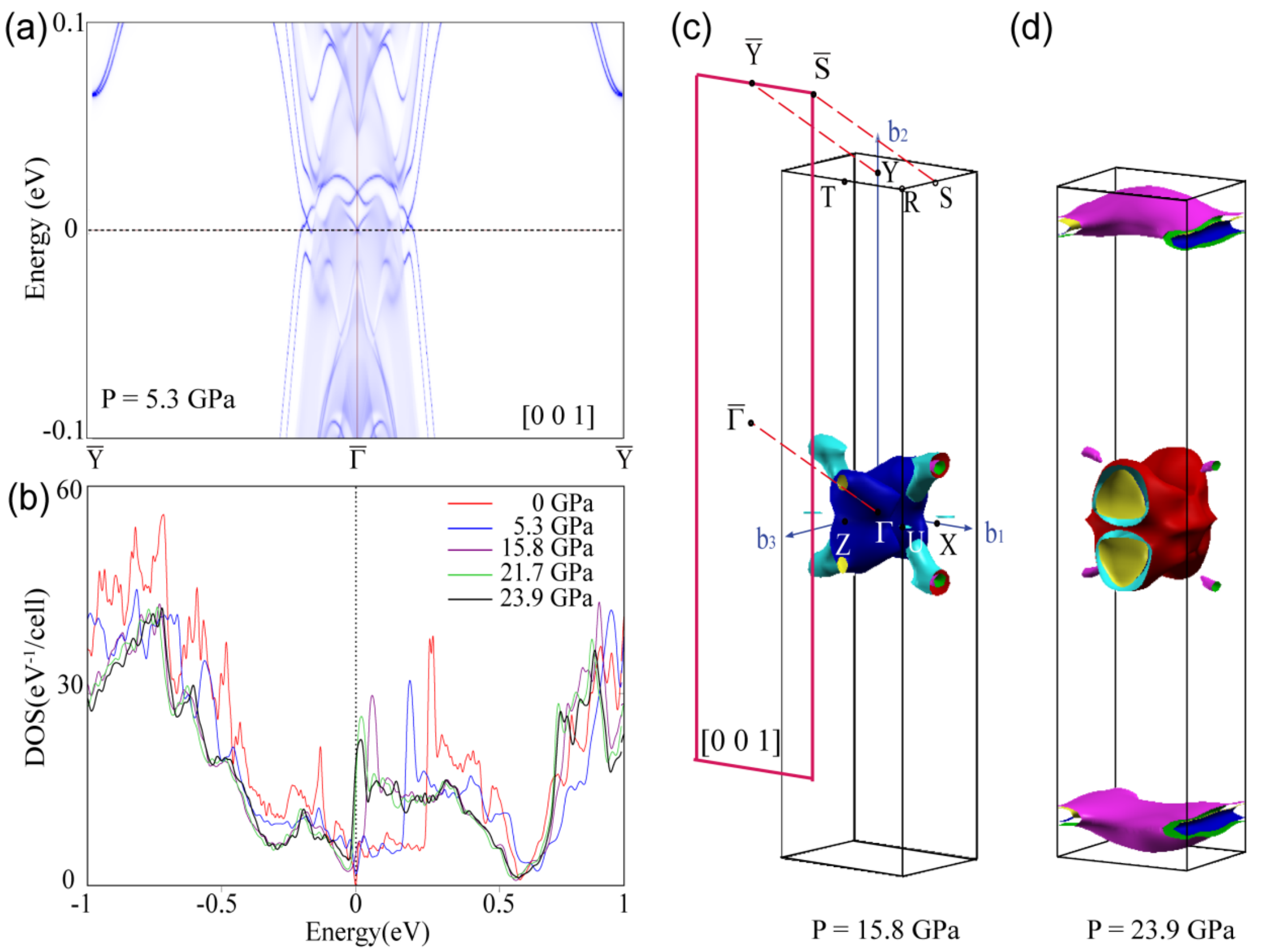

FIG. 5. Pressure evolutions of surface state, density of state and Fermi surfaces of the $Ta_2Ni_3Te_5$. (a) [001] surface state at 5.3 GPa from PBE calculation. (b) Pressure dependent density of state from mBJ calculation. (c,d) [001] surface Brillouin zone and Fermi surfaces for 15.8 GPa and 23.9 GPa, respectively. Middle represents a hole Fermi surface pocket, while flat electron pockets add in top and bottom at 23.9 GPa.

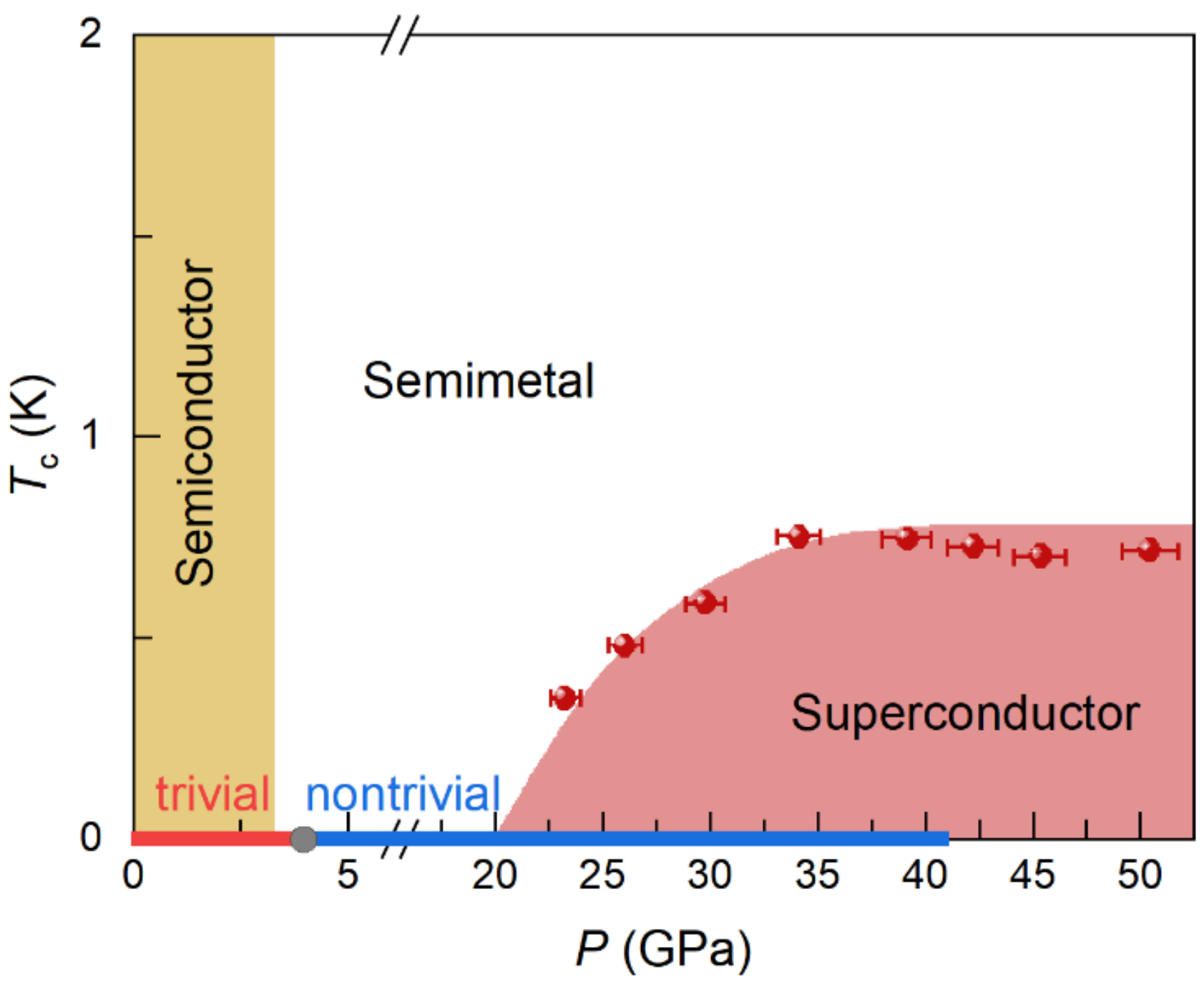

FIG. 6. Pressure-temperature phase diagram of $Ta_2Ni_3Te_5$. The colored areas are guides to the eyes, indicating three distinct conducting states, i.e., semiconductor, semimetal, and superconductor. The red and blue ribbons on the horizontal axis represent the $Z_2$ trivial states and $Z_2$ nontrivial topological states, respectively. The $T_c$ is determined as 90% drop of the normal-state resistivity.

Supplemental Material for

# Pressure-induced nontrivial $Z_2$ band topology and superconductivity in transition metal chalcogenide $Ta_2Ni_3Te_5$

## 1. Methods

### 1.1 Crystal growth and crystallographic characterization

$Ta_2Ni_3Te_5$ single crystals were grown through $I_2$ vapor transport method. Stoichiometric amounts of high-purity Ta, Ni, and Te with a weight of 0.2 g and 2 mg/cm$^3$ of iodide were mixed thoroughly and then sealed in a quartz tube with a low vacuum pressure of ≤ $10^{-3}$ Pa. The tube was heated in a horizontal tube furnace with a temperature gradient of 50°C between 850°C and 800°C for one week. After naturally cooling down to room temperature, many pieces of needle shape and shiny single crystals were obtained. Room-temperature XRD patterns of the single crystal were obtained with Cu $K_\alpha$ radiation ($\lambda$ = 1.5406 Å) using a Rigaku X-ray diffractometer (Miniflex-600). The atomic proportion of the crystal was characterized by energy-dispersive X-ray spectroscopy (EDXS, Helios Nanolab 600i, FEI) with area-scanning mode. The ambient pressure electrical transport measurements were carried out in a physical property measurement system (PPMS, Quantum Design). The dc magnetization was measured in a superconducting quantum interference device (SQUID, Quantum Design).

### 1.2 High-pressure electrical and synchrotron XRD measurements

High-pressure resistances were measured in a nonmagnetic Be-Cu diamond anvil cell using a homemade multifunctional measurement system [20]. High-pressure synchrotron XRD experiments were carried out at room temperature with $Ta_2Ni_3Te_5$ powder crushed from single crystals at pressures up to 50.1 GPa, at the beamline BL15U1 of the Shanghai Synchrotron Radiation Facility (SSRF) (λ = 0.6199 Å). The diamonds with a culet size of 300 $\mu$m and T301 stainless steel gasket were used. Daphne 7373 was used as the transmitting medium. A Mar345 image plate was used to record 2D diffraction patterns. The Dioptas [21] program was used for image integration, and the XRD patterns were fitted using the RIETICA [22] program with the Le Bail method.

The pressure values for all of the above experiments were determined by the ruby fluorescence method at room temperature [23].

### 1.3 First-principles calculations

First-principles calculations were performed with density functional theory (DFT) as implemented in Vienna Abinit Simulation Package (VASP) [24]. The PBE parameterization of generalized gradient approximation to the exchange-correlation functional was employed [25]. We also performed a comparison with the modified Becke–Johnson (mBJ) exchange potentials [26]. Spin-orbit coupling (SOC) for all elements was treated by a second-variation method. A plane-wave basis up 700 eV and 4x12x3 -centered K mesh were employed in the whole calculations under pressure up to 41.1 GPa. The lattice constants under various pressure were used from experimental high-pressure angle-dispersive synchrotron XRD data and the atomic coordinates were fully relaxed until the force on each atom was less than1 meV/A and internal stress was less than 0.1 kbar. The DFT band structures were fitted to a tight-binding (TB) model Hamiltonian using the maximally localized Wannier function (MLWF) method [27]. The $Z_2$ topological indices were calculated using the parity-check method [28], as we did in our previous studies [29-30]. The Fermi surfaces and surface states were calculated using the TB Hamiltonian after being symmetrized with full crystal symmetry [31].

## 2. Synchrotron XRD under pressure

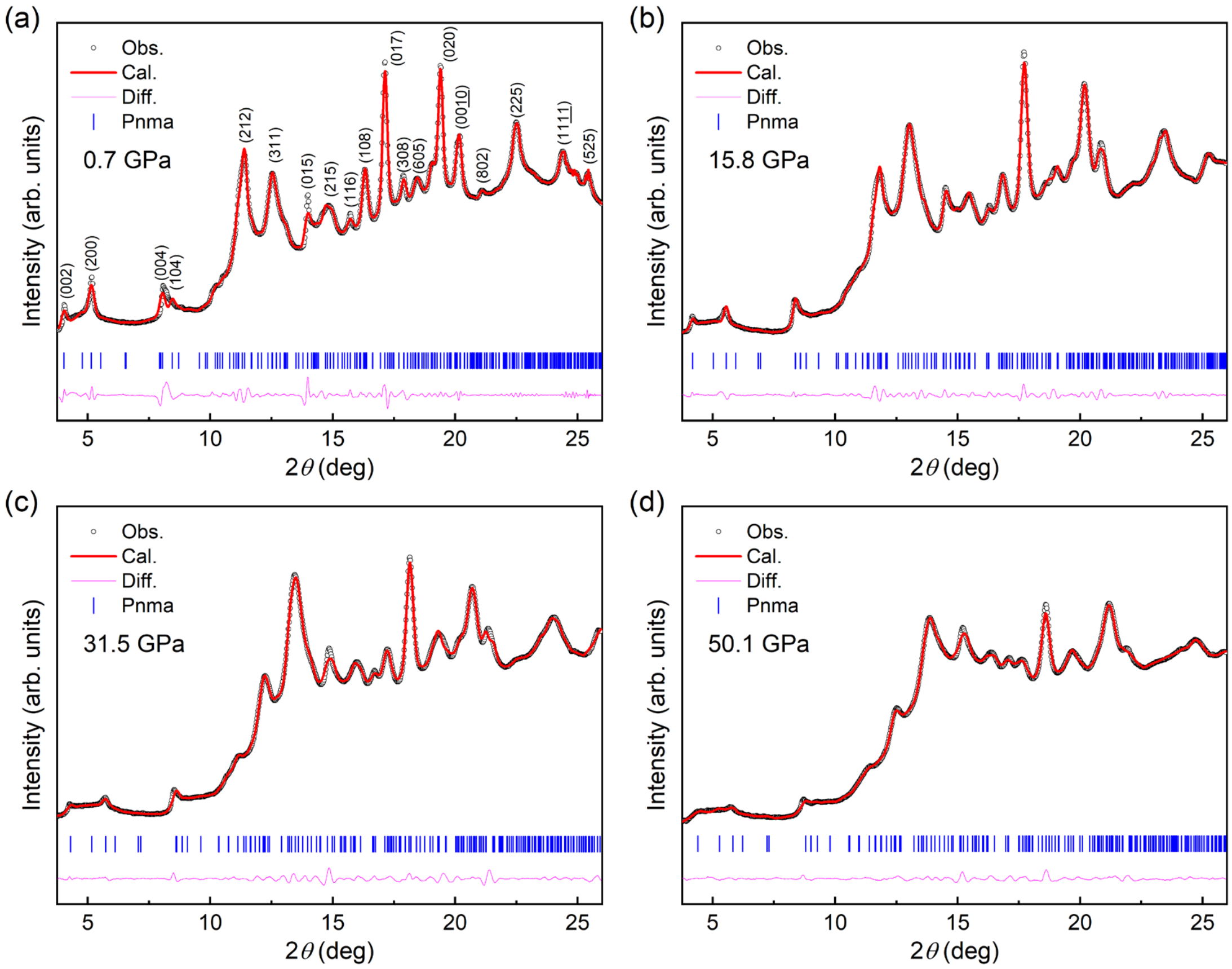

FIG. S1. Representative Le Bail analysis of the XRD patterns at 0.7, 15.8, 31.5, and 50.1 GPa. The open circles and solid lines represent the observed and calculated data, respectively. The solid lines at the bottom denote the residual intensities. The vertical bars indicate the Bragg peak positions with orthorhombic *Pnma* phase.

## 3. Bands structure along Y-Γ-Z under pressure

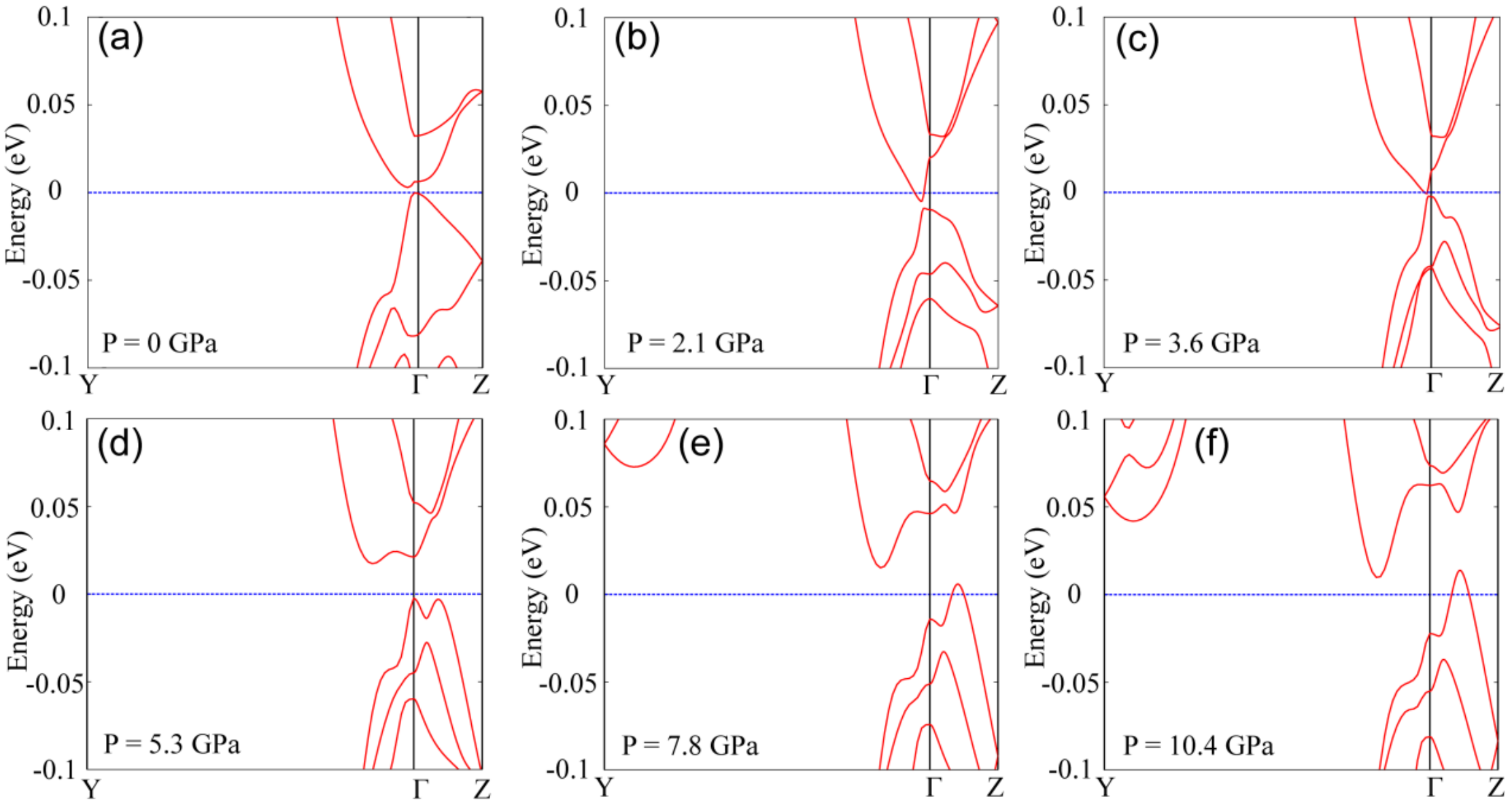

FIG. S2. The evolution of bands structure along Y-Γ-Z from PBE calculation.

## 4. Phase diagram of the direct band gap along Y-Γ versus pressure

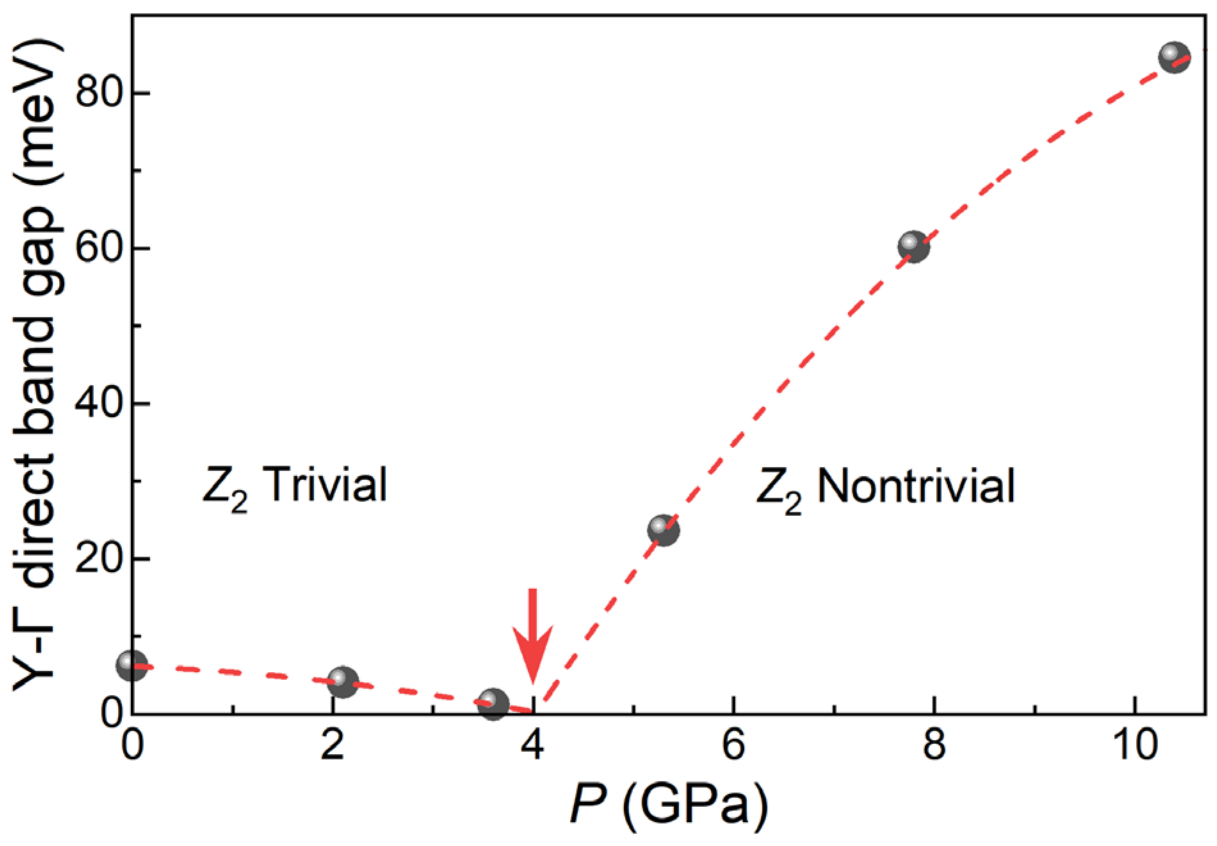

FIG. S3. The direct band gap along Y-Γ as a function of the pressure from PBE calculation. The red arrow denotes the critical pressure value at 4.0 GPa, when the direct band gap along Y-Γ is closed. The low-pressure phase is a $Z_2$ trivial phase, while the high-pressure phase is a $Z_2$ nontrivial phase.

## 5. Density of states at the Fermi level

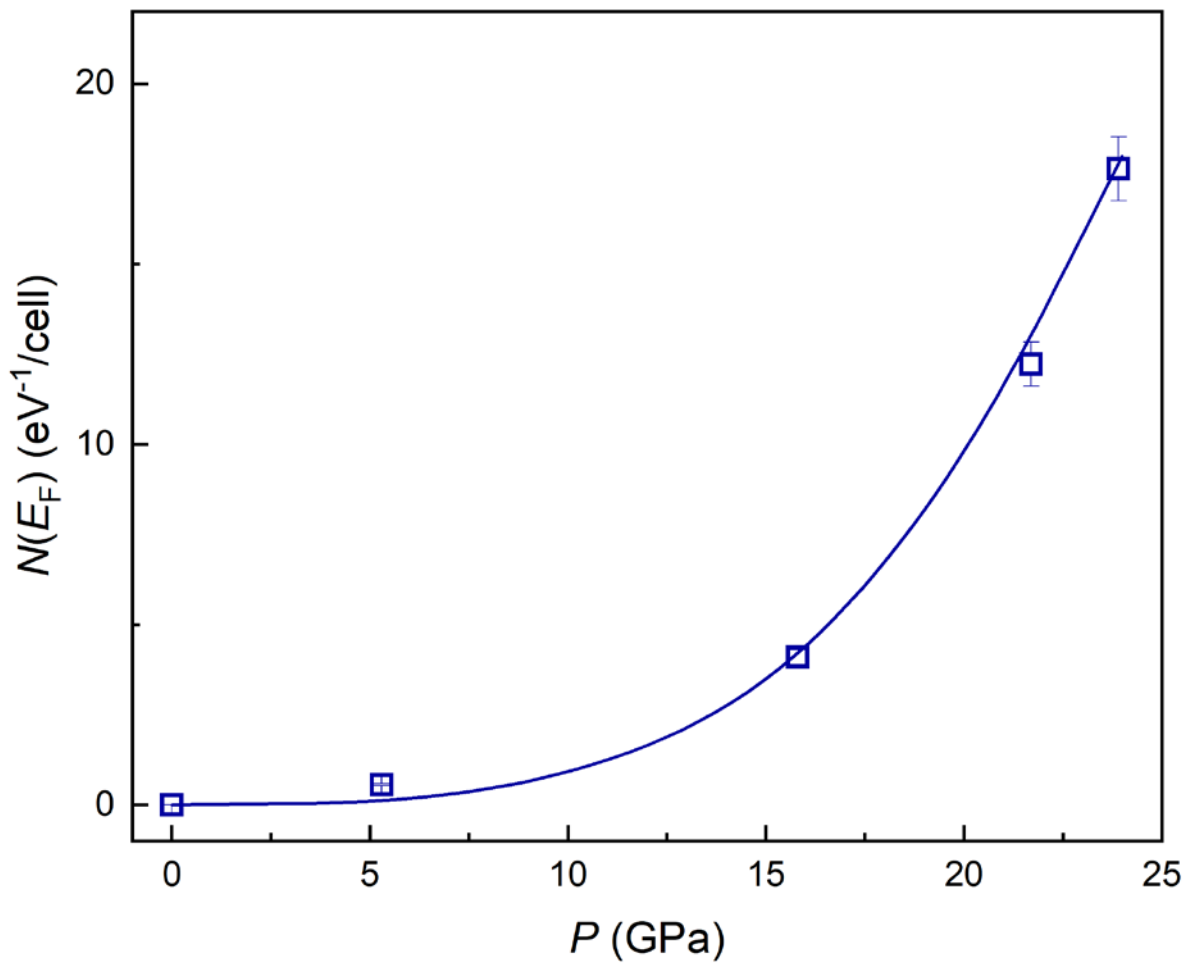

FIG. S4. The calculated density of states at the Fermi level as a function of pressure.

## 6. The $Z_2$ index with the parity under pressure

Table S1. The $Z_2$ index with the parity at the time-reversal invariant momentum (TRIM) points multiplied up to the highest valence band under pressure.

| $P$ (GPa) | Z | Γ | U | X | T | Y | R | S | $Z_2$ |
|---|---|---|---|---|---|---|---|---|---|
| 0 | + | + | + | + | + | + | + | + | (0; 000) |
| 2.1 | + | + | + | + | + | + | + | + | (0; 000) |
| 3.6 | + | + | + | + | + | + | + | + | (0; 000) |
| 5.3 | + | - | + | + | + | + | + | + | (1; 111) |
| 7.8 | + | - | + | + | + | + | + | + | (1; 111) |
| 10.4 | + | - | + | + | + | + | + | + | (1; 111) |
| 15.8 | + | - | + | + | + | + | + | + | (1; 111) |
| 23.9 | + | - | + | + | + | + | + | + | (1; 111) |
| 28.9 | + | - | + | + | + | + | + | + | (1; 111) |
| 34.1 | + | - | + | + | + | + | + | + | (1; 111) |
| 41.1 | + | - | + | + | + | + | + | + | (1; 111) |